# Combination of Convolutional Neural Network and Gated Recurrent Unit for Energy Aware Resource Allocation

Zeinab Khodaverdian
*Department of Computer Engineering, Ayandegan Institute of Higher Education, Tonekabon, Iran*
*Zeinabkhodavardian@aihe.ac.ir*

Hossein Sadr [*]
*Department of Computer Engineering, Rasht Branch, Islamic Azad University, Rasht, Iran [†]*
*Sadr@qiau.ac.ir*

Seyed Ahmad Edalatpanah
*Department of Applied Mathematics, Ayandegan Institute of Higher Education, Tonekabon, Iran*
*Saedalatpanah@aihe.ac.ir*

Mojdeh Nazari Solimandarabi
*Department of Computer Engineering, Qazvin Branch, Islamic Azad University, Qazvin, Iran*
*Mojdeh.nazari@qiau.ac.ir*

**Abstract:** Cloud computing service models have experienced rapid growth and inefficient resource usage is known as one of the greatest causes of high energy consumption in cloud data centers. Resource allocation in cloud data centers aiming to reduce energy consumption has been conducted using live migration of Virtual Machines (VMs) and their consolidation into the small number of Physical Machines (PMs). However, the selection of the appropriate VM for migration is an important challenge. To solve this issue, VMs can be classified according to the pattern of user requests into sensitive or insensitive classes to latency, and thereafter suitable VMs can be selected for migration. In this paper, the combination of Convolution Neural Network (CNN) and Gated Recurrent Unit (GRU) is utilized for the classification of VMs in the Microsoft Azure dataset. Due to the fact the majority of VMs in this dataset are labeled as insensitive to latency, migration of more VMs in this group not only reduces energy consumption but also decreases the violation of Service Level Agreements (SLA). Based on the empirical results, the proposed model obtained an accuracy of 95.18which clearly demonstrates the superiority of our proposed model compared to other existing models.

[*] Sadr@qiau.ac.ir
[†] Department of Computer Engineering, Rasht Branch, Islamic Azad University, Rasht, Iran

## 1. INTRODUCTION

Cloud computing attracted great attention in both industry and research communities for the sake of its ubiquitous, elasticity and economic services [1]. Today, with massive growth of cloud computing in recent years, Service Level Agreement (SLA) and dynamic resource scaling for better services is of great importance [2]. Due to the increasing demand for cloud-based applications, the effective use of cloud resources based on user requests and meeting the level of the service agreement between service providers and consumers have become more difficult [3]. The unpredictable nature of workload also complicates resource allocation in cloud data centers [4]. The lack of optimal allocation not only leads to inefficient use of resources but also yields to excessive energy consumption in cloud data centers [5, 6]. The first class concern of cloud providers is power management for both reducing their total cost of ownership and green computing objectives. Therefore, resource allocation plans in cloud data centers, aiming to reduce energy consumption, include identifying and supplying appropriate resources for workload scheduling and increasing efficiency by providing the maximum utilization of resources [7-9]. Generally, energy management techniques in data centers are classified into two groups of dynamic and static techniques which are illustrated in Figure 1.

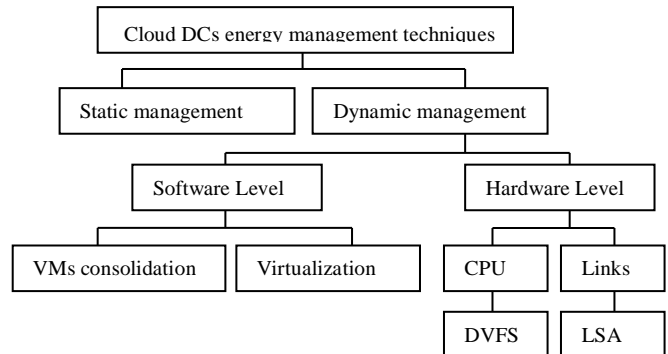

Figure 1. Classification of energy management techniques in cloud data centers

Based on Figure 1, static energy management includes optimization methods in designing step that use efficient hardware components to reduce energy consumption. Despite the efficiency of static energy management techniques, energy consumption is still growing due to the remarkable needs of cloud resources, and dynamic energy management is therefore required [3, 10].

Dynamic techniques are generally performed based on the number of required resources or any other dynamic feature of the system over a period of time. Experiencing variable loading by a system while it is working can be considered as a potential reason that makes dynamic management possible [11]. The ability to infer and predict the future state of the system and make appropriate decisions based on it can be mentioned as another potential reason which is implemented at both hardware and software levels. At the hardware level for instance, by changing the voltage and frequency levels in the processor (DVFS) and using dynamic routing protocols such as Link State Adaptation (LSA) at the network level, the overall energy consumption is dynamically reduced [12]. At the software level, energy consumption is managed by virtualizing on a single server and integrating virtual machines into multiple servers. Virtual machine integration provides optimal performance by supporting the relocation of virtual machines between physical servers and the possibility of live migration [13, 14]. While virtual machines do not use all the provided resources, they can be integrated into a smaller number of physical servers and idle servers can be then disabled to reduce energy consumption in cloud data centers [15, 16].

Accordingly, choosing a virtual machine to migrate toward the minimum physical resources is an important challenge. Given that the dynamic integration of virtual machines is based on live migration, the migration of virtual machines dedicated to Delay-sensitive (Interactive) applications (it refers to virtual machine allocations to programs that require low response times and high performance) can lead to the violation of service level agreements [17-19]. On the other hand, the migration of more virtual machines that are related to Delay-Insensitive applications (generally for programs that are not sensitive to interference), not only leads to energy consumption reduction but also reduces violations of service level agreements [20]. In this regard, virtual machines can be classified based on the user demand and consumption pattern in workload classes (Delay-sensitive or Delay-insensitive classes) and therefore migration candidate virtual machines can be selected based on user demands and increased system efficiency [21, 22]. The imbalance number of samples in Delay-sensitive or Delay-insensitive classes in virtual machine classification can be considered as another challenge. As a matter of fact, the number of virtual machines that belong to the Delay-insensitive class is more than the number of virtual machines that belong to Delay-sensitive classes which may affect the classification performance [23].

In recent years, deep learning techniques have emerged as an effective solution for virtual machine classification [24, 25]. Deep learning is a subset of machine learning that can learn features without human intervention [26,27]. deep learning is in a hierarchy of increasing complexity and abstraction and therefore can extract nonlinear patterns from the virtual machine workload. In fact, deep learning can extract multiple correlations between virtual machines based on previous workloads and predict their future workloads with high accuracy [28]. Virtual machine load prediction helps to make decisions about capacity planning and use suitable location and virtual machine migration. Consequently, the following solutions are presented in this paper to overcome the mentioned issues.

- Due to the fact that a large number of samples are in the latency-insensitive class, the distribution of samples is unbalanced. To this end, SMOTE method was utilized to deal with the data imbalance problem in this paper.
- The combination of CNN and GRU was utilized to perform the virtual machine load classification.
- Based on the empirical result, employing the combination of CNN and GRU increased the classification accuracy. Moreover, dropout was used as a generalization technique in our proposed model which not only reduced overfitting but also increased the robustness of our model.

- Using the fewer number of layers in the proposed model also decreased complexity and computational cost.

The remainder of this paper is organized as follows. Section 2 includes the literature review. The proposed model for virtual machine classification is described in Section 3. Section 4 presents the experiments and empirical results. Conclusion and future research directions are also provided in Section 5.

## 2. RELATED WORK

Common virtual machine selection techniques include random selection, minimum utilization, maximum correlation, minimum migration time,  utilization slope, and machine learning [29]. The virtual machine selection techniques are depicted in Figure 2.

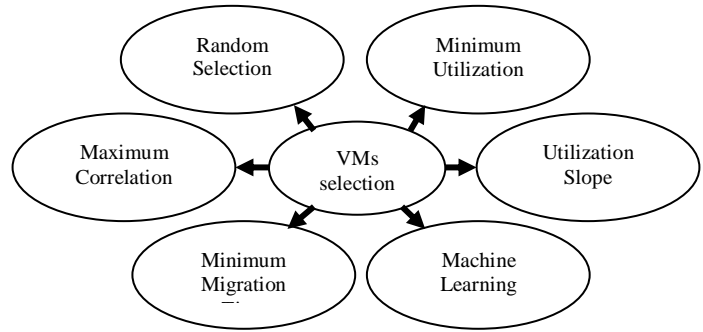

Figure 2. Virtual machine selection techniques

In random selection, a virtual machine is randomly selected for migration. Minimum migration time is also another factor for choosing the virtual machine which is equal to the amount of memory used by the virtual machine divided by the bandwidth of the host (physical server where the virtual machine is located) [30]. In another method, virtual machines that have the lowest processor utilization can be bee chosen for migration. In the maximum correlation technique, the virtual machine is selected based on the highest correlation (in terms of processor efficiency) [31]. Utilization slope techniques also employ the utilization history of the virtual machines and compute the slope between the utilization points in each particular time period for each virtual machine and then the virtual machine with a higher total utilization slope is selected for migration.

### 2.1. *Machine learning-based virtual machine selection*

Due to the fact that choosing the best virtual machine has a remarkable impact on energy management, using intelligent machine learning methods can yield superior performance in this field which can be done by predicting the workload of the virtual machine. Prediction can be defined as the determination of the value of a dependent variable in terms of the values of the independent variable and regression is known as the most important method of numerical prediction. Rafael Moreno et al. [32]  proposed a novel predictive auto-scaling mechanism based on machine learning techniques for time series forecasting and queuing theory. The new mechanism aims to accurately predict the processing load of a distributed server and estimate the appropriate number of resources that must be provisioned in order to optimize the service response time and fulfill the SLA contracted by the user, while attenuating resource over-provisioning in order to reduce energy consumption and infrastructure costs. The results show that the proposed model obtains a better forecasting accuracy than other classical models, and makes a resource allocation closer to the optimal case. Linear regression, ridge regression, lasso regression, and support vector regression were the techniques that were tested on the Microsoft Azure dataset by Striki et al. [33]. Shaw et al. [34] also examined ensemble learning including support vector machine, neural network, and logistic regression to classify virtual machines in the Microsoft Azure dataset. They randomly reduced the number of samples in the majority class to deal with the data imbalance problem in the dataset. Based on their obtained results, using the boosting method for support vector machine can surprisingly enhance the overall performance. Following a similar line of research, Sharma et al. [35] utilized the boosting method with a decision tree and obtained an accuracy of about 74% for virtual machine classification. Witanto et al. [36] trained a neural network on the PlantLab dataset in 100 epochs. They obtained the accuracies of 74.92% and 75.05% respectively on train and test datasets. Zhang et al. [37] examined the efficiency of the Recurrent Neural Network (RNN) compared with regression-based models for virtual machine selection on Google dataset. Based on the experimental results, RNN presented higher accuracy in comparison to the regression-based models. Although RNN is suitable for processing time series data, they are suffering from long time dependency problem. To overcome this issue, Zhu et al. [38] utilized the Long Short Term Memory (LSTM) network that could preserve long time dependencies using memory cells. Their proposed model presented higher performance compared to both RNN and GRU. Since recurrent neural network (RNN) is naturally suitable for sequential data analysis, it has been recently used to tackle the problem of workload prediction. However, RNN often performs poorly on learning long-term memory dependencies, and thus cannot make the accurate prediction of workloads. To address these important challenges, Chen et al. [39] proposed a deep Learning based Prediction Algorithm for cloud Workloads. First, a top-sparse auto-encoder (TSA) is designed to effectively extract the essential representations of workloads from the original high-dimensional workload data. Next, thay integrate TSA and gated recurrent unit (GRU) block into RNN to achieve the adaptive and accurate prediction for highly-variable workloads. Using real-world workload traces from Google and Alibaba cloud data centers. Moreover, the performance results show that thair achieves superior prediction accuracy compared to the classic RNN-based and other workload prediction methods for high-dimensional and highly-variable real-world cloud workloads.

Moreover, Aslam et al. [40] examined RNN, LSTM, Boltzmann machine, and CNN on the PlantLab dataset. Ohamey et al. [41] used a hybrid method for predicting virtual machine workload on real data from 1750 virtual machines in a distributed data center. based on the Vector Autoregressive (VAR) model and the Stacked Long Short Term Memory (LSTM) model. In thair proposed method, two metrics are used: CPU and memory usage, the VAR model is used to filter the linear interdependencies among the multivariate time series, and the stacked LSTM model to capture nonlinear trends in the residuals computed from the VAR model. The proposed hybrid model is compared with other hybrid predictive models: the ARMLP model, the RNN-GRU model and the ARIMA-LSTM model. Results of experiments show superior efficacy of the proposed method over the other hybrid models.Yazdanian and Sharifian [42] proposed a model combined 1D ConvNets and stack of long-short term memory (LSTM) blocks to process long sequence of Google trace data in order to have a precise and light computing prediction of RAM and CPU requests in future timestamps. Experimental results confirms that thair approach while having high accuracy, also not involved in heavy calculations and efficiently works with long sequences. A summary of studies is provided in Table 1.

TABLE 1. A SUMMARY OF STUDIES

| Year | Authors | Dataset | Proposed Model | |
| --- | --- | --- | --- | --- |
| | | | Pre-Processing (Data balancing) | Model |
| 2019 | Rafael Moreno et al. [32] | Complutence datasets | Imbalance dataset | support vector regression |
| 2018 | Striki et al. [33] | Microsoft Azure dataset | Imbalance dataset | Linear regression, ridge regression, lasso regression, and support vector regression |
| 2019 | Shaw et al. [34] | Microsoft Azure dataset | Random under sampling (RUS) | Ensemble learning approach |
| 2017 | Sharma et al. [35] | UCI | Random over sampling (ROS) | Boosted tree |
| 2018 | Witanto et al [36] | PlantLab dataset | Imbalance dataset | Artificial Neural Network |
| 2016 | Zhang et al. [37] | Google dataset | Imbalance dataset | Recurrent Neural Network |
| 2019 | Zhu et al. [38] | Google dataset | Imbalance dataset | Long Short Term Memory |
| 2019 | Chen et al. [39] | Alibaba cloud data and Google dataset | Imbalance dataset | TSA-GRU |
| 2019 | Aslam et al. [40] | PlantLab dataset | Random over sampling (ROS) | RNN, LSTM, Boltzmann machine, and CNN |
| 2019 | Ohamey et al. [41] | 1750 virtual machines in a distributed data center | Imbalance dataset | VAR-LSTM |
| 2018 | Yazdanian and Sharifian [42] | Google dataset | Imbalance dataset | CNN-LSTM |

Considering the mentioned studies that have focused on using intelligent methods to perform virtual machine selection, deep learning-based techniques presented higher efficiency compared to machine learning-based techniques while they do not require feature engineering [43]. Since the workload of virtual machines is collected over a period of time and in a regular sequence, techniques such as RNNs that are able to record long time dependencies have superior performance. On the other hand, RNNs are facing gradient vanishing and exploding problems and cannot extract local features effectively. The convolutional neural networks can perform well in extracting local features but is weak in extracting long-term dependencies. As seen in previous studies, hybrid models performed well and greatly improved the results. The reason for using hybrid models is that models can cover each other's weaknesses and cause synergies. But on the other hand, hybrid models are more complex than other models. In this regard, a shallow deep neural network which is the combination of CNN and GRU is proposed in this paper which aims to overcome the mentioned challenges.

## 3. PROPOSED MODEL: COMBINATION OF CNN AND GRU

The details of the proposed model including pre-processing step and model description are provided in the following of this section. The diagram of the proposed model is also depicted in Figure 3.

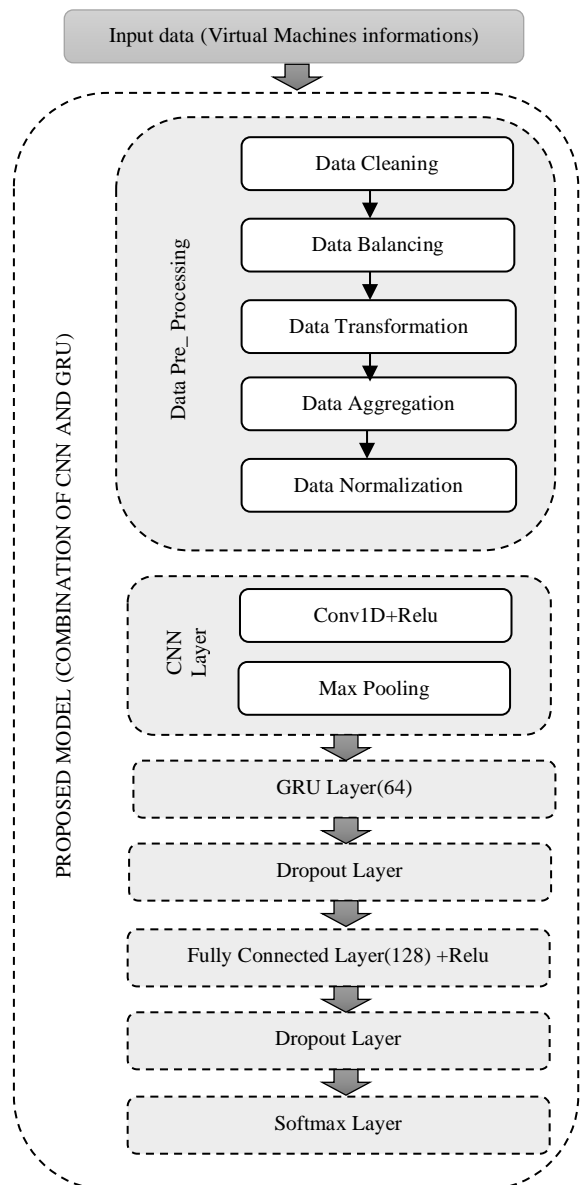

Figure 3. Diagram of the proposed model

### 3.1. *Pre-Processing*

The pre-processing steps of the proposed model are mentioned in the following.

### 3.1.1. *Data cleaning*

Records with unknown labels are eliminated from the dataset.

### 3.1.2. *Data balancing*

While the number of samples is unequally distributed in various classes of a dataset, the class imbalance problem occurs. Classes that have a smaller number of samples compared to other target classes are called minority classes and the others are called majority classes [44]. To extract knowledge from unbalanced datasets, minority classes have great importance and minority class samples must be correctly increased. SMOTE [a] is a method for increasing the minority class samples. In this method, new artificial samples are generated in the neighborhood of existing samples and are then placed in a line. Thereafter, they are connected to the samples of adjacent minority classes. Notably, the characteristics of samples that are in the adjacent classes are not changed, and therefore SMOTE can generate samples that belong to the main distribution. Here, $D_{new}$ is the new generated artificial data which is calculated using the following equation (Eq. 1).

$$D_{new} = D_i + (D\hat{}_l - D_i) \times \delta \tag{1}$$

Where $D_i$ is the number of minority class samples, $D\hat{}_l$ is one of the close neighbors, and $\delta$ is a random number between zero and one. Generating new samples in the neighborhood of minority class sample can be considered as one of the most important advantages of SMOTE method compared to other resampling techniques. On the other hand, this method is also easier and less complex than other data balancing techniques, such as cost-sensitive methods.

### 3.1.3. *Feature transformation*

Label encoding is used in our proposed model to transform nominal attributes into numeric ones to become interpretable for neural networks. Label encoding considers a number starting from zero to n-1 for each sample with nominal properties [43,45]. The reason behind choosing this method is that it does not change the dimensionality of the data.

### 3.1.4. *Data Aggregation*

It includes operations that lead to the generation of new features via combining two or more features. The new features must be able to represent the information of the dataset more efficiently and completely compared to the original features. Using data aggregation in the proposed model not only reduces dimensionality but also increases the value of features and data stability.

---

[a] Synthetic Minority Over-sampling Technique (SMOTE)

### 3.1.5. *Normalization*

Max–Min normalization method is used in the proposed model to normalize the input data. This method performs a linear transformation on the original data and maintains the correlation between them. Due to the fact that the correlation between data and the relation between independent variables are effective in prediction, this normalization method is utilized (Eq. 2).

$$x\square = \frac{x - Min(A)}{Max(A) - Min(A)} \tag{2}$$

Where *Min(A)* and *Max(A)* respectively refer to the minimum and maximum value of a feature and x is equal to the current value of the feature.

## 3.2. *CNN layer*

The input data including the virtual machine loading information enters the CNN after the pre-processing steps. The CNN consists of a convolution layer and a max-pooling layer. The convolution layer contains some filters. By repeatedly applying the convolutional filter on the input matrix, the feature maps are obtained and the learning process includes updating the filter elements which are the weights of the neural network [46- 48]. The obtained feature maps must then pass through the nonlinear RelU activation function which makes the network able to learn the nonlinear patterns in the data. Whereas the size of the feature maps is not constant and is related to the size of input matrix and filters, pooling function is required to induce a fixed-length vector which is considered as feature selection. Therefore, the max-pooling layer is used after the convolution layer to reduce the size of feature maps and which can also lead to the selection of immutable and more valuable features [49, 50]. The reason behind using CNN is that it can extract local features [51-53]. Considering that the workloads of virtual machines are collected in a period of time and a regular sequence, the obtained features are then entered into a GRU to extract long time dependencies.

## 3.3. *GRU layer*

Due to the fact that CNNs are not able to extract long time dependencies, the output of the CNN layer is then fed to a GRU. Although increasing the number of layers in CNN can also yield to extract long time dependencies, it is costly and increases the computational cost [54,55]. On the contrary, GRU is able to overcome this issue. On the other hand, GRU includes reset and update gates that can help to overcome gradient vanishing and exploding problems in RNNs and it also has a fewer number of training parameters compared to LSTM. Reset and update gates in GRU are two vectors [56] that create a very precise control mechanism using the Sigmoid activation function and decide how much of the previous memory is related to the current input [57]. The architecture of GRU is depicted in Figure 4.

Based on Figure 4, the input of the current state is indicated by c<t-1>, which is also the output of the previous state. The output of the current state, which is also the input of the next state, is indicated by c <t>. x <t> and y^ <t> also refer to the input and output of current state. Γu and Γr also respectively state the update gate and the reset gate that are calculated using (Eq. 3) and (Eq. 4). c~<t> is the hidden state which is computed using (Eq. 5) similar to RNN and finally c <t> is computed using (Eq. 6).

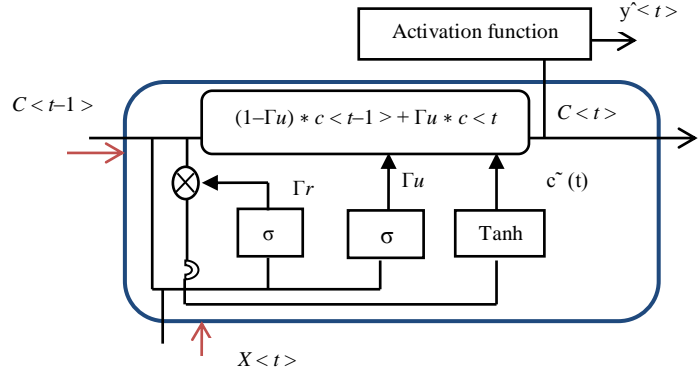

Figure 4. The architecture of GRU

$$\Gamma u = \sigma(Wu[c\langle t-1 \rangle, x\langle t \rangle] \ + \ bu) \tag{3}$$

$$\Gamma r = \sigma(Wr[c\langle t-1 \rangle, x\langle t \rangle] \ + \ br) \tag{4}$$

$$c\tilde{}\ <t> = tanh(Wc[\Gamma r * c\langle t-1 \rangle, x\langle t \rangle] \ + \ bc) \tag{5}$$

$$c<t> = (1-\Gamma u) \ * \ c<t-1> \ + \Gamma u \ * \ c\tilde{}<t> \tag{6}$$

Where $bc$ ، $br$ ، $bu$ refer to bias, $\sigma$ is the sigmoid activation function and $Wu$ ، $Wr$ ، $Wc$ are the weight matrices.

### 3.4. *Fully connected(FC) layer*

Fully Connected layers in a neural networks are those layers where all the inputs from one layer are connected to every activation unit of the next layer. In most popular machine learning models, the last few layers are full connected layers which compiles the data extracted by previous layers to form the final output [58-60].

Fully connected(FC) layer is designed to handle vector info efficiently.The depth of model should be carefully determined. We applied one layers of Fully connected layers here, but it can be adjusted by serviceprovider to achieve the balance between the complexity of targetmodel (Better detection accuracy) .

### 3.5. *Dropout layer*

The dropout layer is used to prevent over-fitting. In fact, dropout is a regularization technique that is used in neural networks to reduce repetitive learning between neurons. To this end, some neurons are randomly ignored along with the training process [61- 63].

### 3.6. *Classification layer (Soft max)*

Softmax is used in the final layer to perform classification. Softmax is a nonlinear activation function that is used in the output layer of a neural network to conduct final classification [64]. Softmax is computed using Eq. 7 and the output values are normalized in a way that the summation of values is equal to one.

$$P(Y = k|X = x_i) = \frac{e^{sk}}{\sum_j e^{sj}} \tag{7}$$

Where s is the k-dimensional input vector and $esk$ is the standard exponential function that is applied to each element of the input vector. The denominator of the fraction ensures that all output values are in the range of zero and one. In the following, the score of the correct class must be maximized. To this end, the logarithm in Eq. 8 must be maximized, and to minimize the cost function, the logarithm must be multiplied by a negative.

$$Li = -log \frac{e^{syi}}{\sum_j e^{sj}} \tag{8}$$

## 4. EXPERIMENTAL AND RESULTS

In this section, datasets, evaluation metrics, software and hardware requirements are discussed. Experiments and empirical results are also provided in the following.

### 4.1. *Dataset*

The dataset used in our experiment includes information about all Microsoft Azure virtual machines collected in 3 months from November 16, 2016, to February 16, 2017 [9]. The description of the Microsoft Azure virtual machines dataset is provided in Table 2.

TABLE 2. DETAILS OF MICROSOFT AZURE DATASET

| Dataset characteristics | Description |
|---|---|
| Dataset storage type | Record (matrix) |
| Dataset dimensionality | Two-dimensional |
| Number of rows (number of samples) | 1048578 |
| Number of columns (number of features) | 10 feature columns and one label column |
| Feature type | Nominal and numerical |
| Missing data | No |
| Unknown label | Yes |
| Number of labels | 2 |
| Distribution of samples in each class | Unbalanced |
| Storage format | CSV |

This dataset attributes also include virtual machine ID, Subscription and deployment ID, virtual machine creation and elimination timer, minimum, average, and maximum CPU usage, virtual machine virtual core counter, and virtual machine memory. Processor usage is measured every 5 minutes. Virtual machines in this dataset are divided into two classes Delay-sensitive (Interactive) or Delay-insensitive.

### 4.2. *Evaluation metrics*

Accuracy metric (Eq. 9), Precision metric (Eq. 10) and Recall metric (Eq. 11) are used to evaluate the proposed model where TP and TN respectively refer to the positive and negative samples that are correctly classified. FP and FN also respectively refer to positive and negative examples that are incorrectly classified.

$$Accuracy = \frac{TP+TN}{TP+FP+TN+FN} \tag{9}$$

$$Precision = \frac{TP}{TP+FP} \tag{10}$$

$$Recall = \frac{TP}{TP+FN} \tag{11}$$

### 4.3. *Experiment description*

Experiments were conducted on the hardware including Intel processor Core ™ i5 cores with a speed of 1.60 GHz. Google Colab laboratory (Colaboratory) was also used as to simulate the proposed model. Experiments were started with data pre-processing. The samples with unknown labels were first removed from the dataset. The distribution of samples in different classes is depicted in Figure 5.

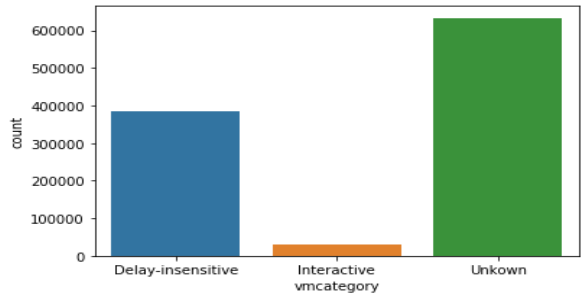

Figure 5. Distribution of samples in different classes

The total number of samples was about 1048576 and only 415238 of them were remained after removing the samples with unknown labels. Due to the fact that the label encoding method was used for feature transformation in this paper, the column including dataset's labels was also replaced with zero and one. The zero indicates samples that are not sensitive to latency and one indicates the samples that are sensitive to latency. Sample distribution after removing samples with unknown labels is illustrated in Figure 6.

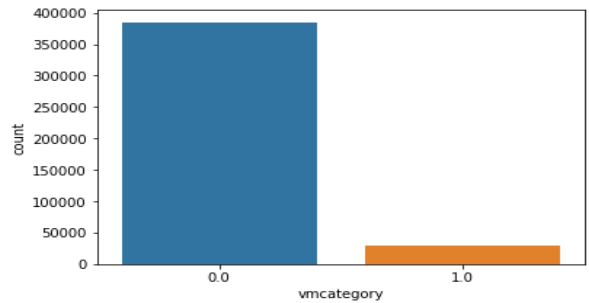

Figure 6. Distribution of samples in different classes after removing samples with unknown labels

SMOTE method was then utilized to balance the dataset. Therefore, by increasing the number of samples in latency-sensitive class, the total number of samples in the dataset was about 770172. In the data aggregation step, the number of features was reduced to 9 from 11.

We get the VM lifetime in hours by subtracting the "vmcreated" value from "vmdeleted" value and dividing by 3600. We then, populate a new column called "core-hour" by multiplying this lifetime with the vmcorecount value. New features will be added to the dataset and previous features will be removed. This creates more correlation between features. The correlation diagram before data aggregation is shown in Figure 7 and after data aggregation in Figure 8.

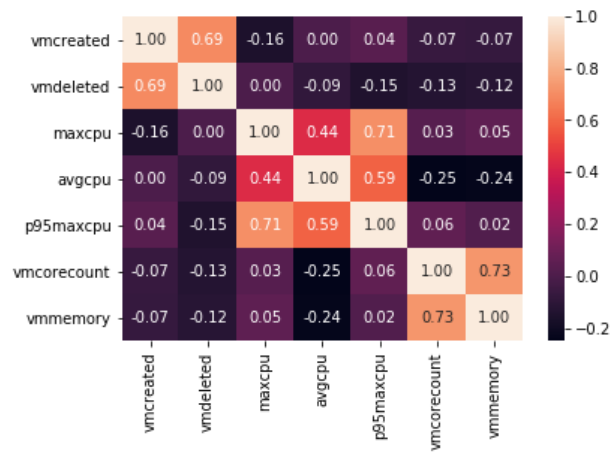

Figure 7. correlation diagram before data aggregation

While pre-processing is finished, the model training is started. Model tuning and hyperparameters' values are mentioned in the following section.

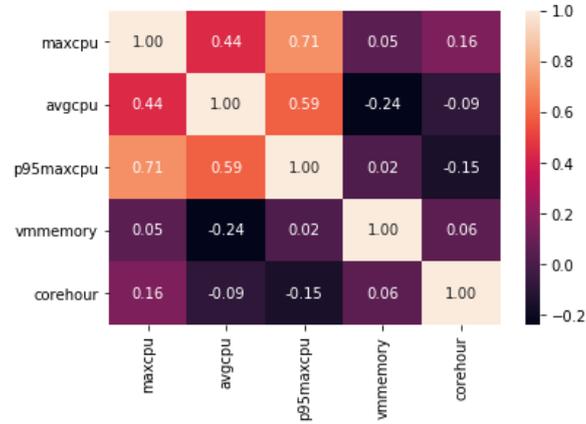

Figure 8. correlation diagram after data aggregation

### 4.4. *Training and hyper parameters*

After normalization, 70% of the samples were used as the training set while about 10% and 20% of the remained samples were respectively used as the validation and test sets. The training set was first entered into the model and the initial weights were randomly set. In the forward pass, the weights were multiplied by each neuron in the input data and the bias values were then added. During the training process, parameters' values, including weight and bias, were updated in the back propagation to obtain an optimal value. To this end, the difference between the actual values and the predicted values was considered as the loss value in each epoch and the gradient of the loss function was calculated in relation to the weight. The weights were then updated in the next epoch to minimize the loss value. The mentioned process was repeated until the loss value was converged. The values of the hyper parameters are depicted in Table 3.

Table 3. Hyperparameters values Table

| Hyper parameters | Values |
|---|---|
| Neural network Type | CNN and GRU |
| Classification function | Softmax |
| Activation function | Relu |
| Number of filters | 64 |
| Filter size | 3 |
| Pooling size | 2 |
| Bach size | 64 |
| Dropout rate | 0.4 |
| Number of epochs | 100 |
| Optimizer | ADAM |
| Learning rate | 0.01 |
| Loss function | Cross-Entropy |

#### 4.5. *Results*

The empirical results are reported in this section. The proposed model was trained in 100 epochs while the number of samples in the training set, validation set, and test set were respectively about 519866, 57763, and 192545. The training process on test and validation sets of the Microsoft Azure dataset is depicted in Figure 9.

The training process is finished after 100 epochs and then the trained model is evaluated. To provide comprehensive analysis, the proposed model was trained 5 times and the maximum, minimum values besides mean and standard deviation of 5 runs are reported. The obtained results are provided in the following.

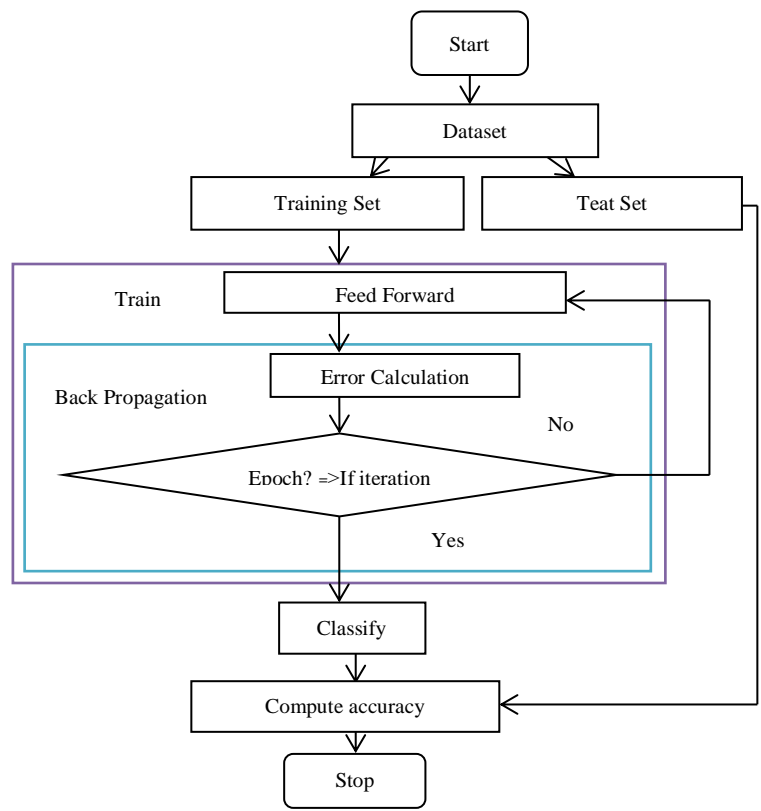

Figure 9. Flowchart of the training process

#### 4.5.1. *Experimental results*

In order to provide a better analysis of the performance of the proposed model, the plot of classification accuracy and loss values per epoch over the training and validation sets are depicted in Figure 10 and Figure 11. As it is clear, the proposed model is converged after 60 epochs on the training set and the highest accuracy is obtained after 87 epochs. The curve of the validation set also illustrates the generalization of the proposed model.

As matter of fact, using dropout not only prevents overfitting but also improves the generalization of the proposed model.

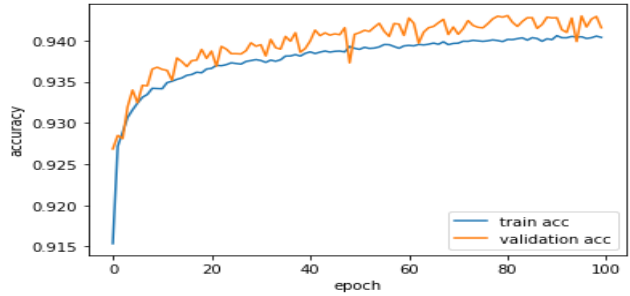

Figure 10. classification accuracy per epoch over the training and validation sets

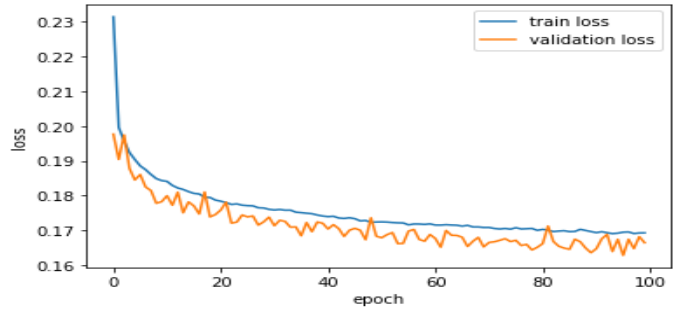

Figure 11. Loss values per epoch over the training and validation sets

The accuracy of the proposed model compared to other existing models is presented in Table 4. Notably, RUS and ROS techniques besides SMOTE method were used in our experiment to balance the data in every 5 runs of the model.

Table 4. ACCURACY (%) COMPARISON

| Model | Accuracy (%) | | |
|---|---|---|---|
| LR [12] | 70.44 | | |
| SVM[13] | 72.34 | | |
| ANN[31] | 84.06 | | |
| CNN [17] | 91.89 | | |
| RNN [15] | 92.88 | | |
| CNN-LSTM+ SMOTE | 95.11 | | |
| | Min | Mean(std) | Max |
| CNN-GRU+ SMOTE | **95.01** | **95.18(0.13)** | **95.23** |
| CNN-GRU+RUS | 93.27 | 93.31(0.39) | 94.61 |
| CNN-GRU+ROS | 94.12 | 94.34(0.17) | 94.98 |

In order to provide a better analysis of the performance of the proposed model, number of epochs, number of Trainable parameters and Training Time for the proposed model and other models are depicted in Table 5.

Table 5. COMPARISON OF THE NUMBER OF EPOCHS, NUMBER OF TRAINABLE PARAMETERS AND TRAINING TIME FOR THE PROPOSED MODEL AND OTHER MODELS

| Model | Epoch | Trainable parameters | Training Time(S) |
|---|---|---|---|
| ANN [34] | 100 | 16896 | 588.31 |
| CNN [38] | 100 | 17154 | 946.52 |
| RNN [35] | 100 | 20960 | 967.14 |
| CNN-LSTM+ SMOTE | 100 | 33410 | 1702.43 |
| **CNN-GRU+ SMOTE** | **100** | **25346** | **1577.87** |

The precision of the proposed model compared to other existing models is presented in Table 6 and The recall of the proposed model compared to other existing models is presented in Table 7.

Table 6. PRECISION COMPARISION

| Model | Precision | |
|---|---|---|
| | Interactive | Delay-Insensitive |
| LR [31] | 0.68 | 0.81 |
| SVM[32] | 0.64 | 0.89 |
| ANN[34] | 0.71 | 0.79 |
| CNN [38] | 0.80 | 0.92 |
| RNN [35] | 0.83 | 0.93 |
| CNN-LSTM+ SMOTE | 0.90 | 0.97 |
| **CNN-GRU+ SMOTE** | **0.89** | **0.96** |
| CNN-GRU+RUS | 0.80 | 0.94 |
| CNN-GRU+ROS | 0.87 | 0.91 |

Table 7. RECALL COMPARISION

| Model | Recall | |
| --- | --- | --- |
| | **Interactive** | **Delay-Insensitive** |
| LR [31] | 0.85 | 0.59 |
| SVM[32] | 0.94 | 0.43 |
| ANN[34] | 0.80 | 0.64 |
| CNN [38] | 0.84 | 0.76 |
| RNN [35] | 0.83 | 0.80 |
| CNN-LSTM+ SMOTE | 0.88 | 0.89 |
| **CNN-GRU+ SMOTE** | **0.87** | **0.91** |
| CNN-GRU+RUS | 0.76 | 0.81 |
| CNN-GRU+ROS | 0.82 | 0.88 |

comparison of the accuracy is presented in Figure 12  And comparison of the Precision and recall are presented in Figure 13 and Figure 14.

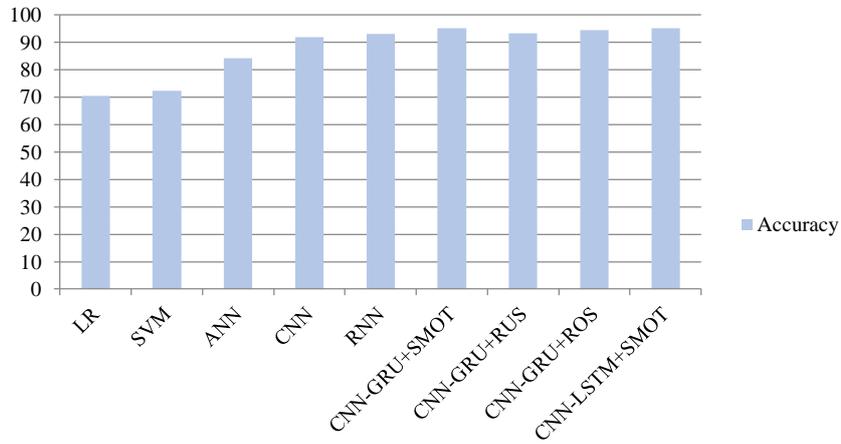

Figure 12. comparison of the accuracy

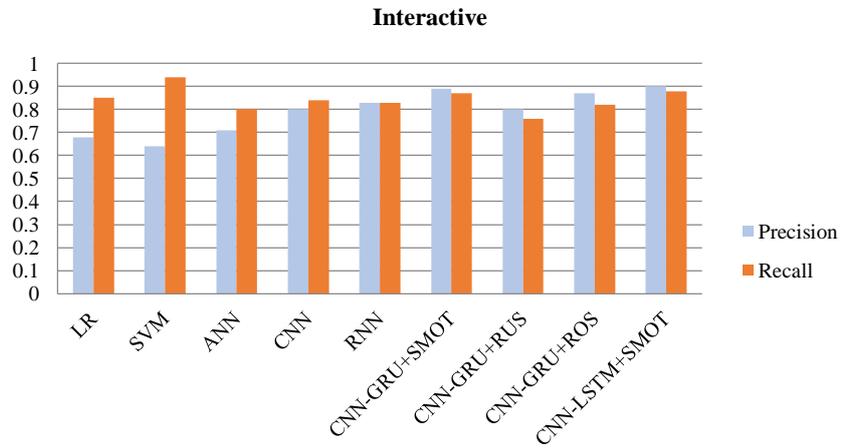

Figure 13. comparison of Precision and Recall in Interactive class

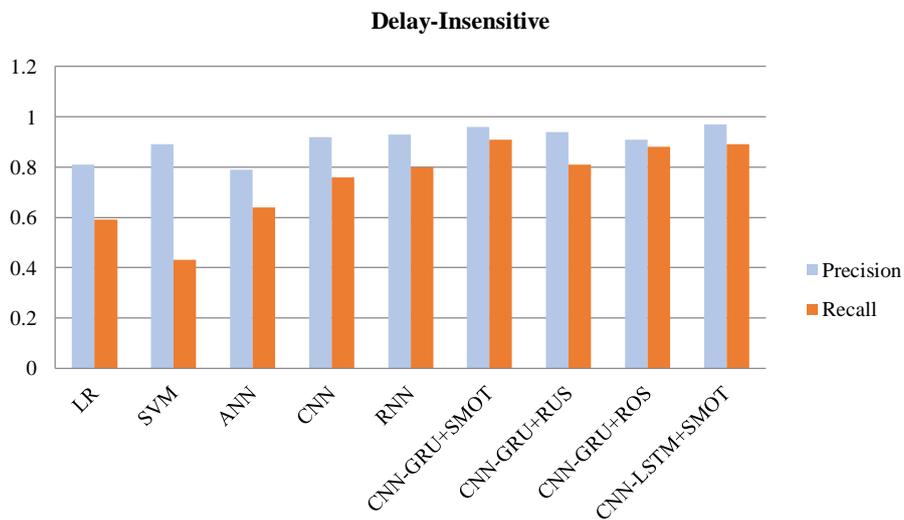

Figure 14. comparison of Precision and Recall in Delay-Insensitive class

### 4.6. *Discussion*

Considering the accuracy of the proposed model in comparison to other existing models, it is clear that the proposed model has superior performance. In this regard, it can be stated that:

1) Using SMOTE method to overcome the data imbalance distribution of the dataset surprisingly increased the classification accuracy because the accuracy did not deviate to the majority class in the balance data distribution. Moreover, SMOTE method presented higher performance compared to resampling techniques. In resampling techniques, existing samples are copied or deleted. In random under sampling (RUS), removing the majority class samples may lead to the elimination of valuable samples. In random over sampling (ROS), the number of samples in the minority class is increased by copying or reusing samples which not only increases the training time but also requires monitoring and adjusting the samples. However, using SMOTE method, new samples are generated in the vicinity of the previous samples.

2) Deep learning-based methods presented superior performance compared to machine learning-based methods for virtual machine load prediction. Due to the fact that deep neural networks are able to extract features automatically, they can extract multiple correlations between virtual machines based on the previous load and therefore predict their future load with high accuracy.

3) Employing the combination of CNN and GRU has increased the classification accuracy. As a matter of fact, using the combination of these models helps us to make use of their advantages as well as overcoming their pitfalls.

4) Deep neural networks commonly include numerous hidden layers besides input and output layer that can increase their performance. In contrast, shallow networks utilize a lower number of the hidden layers which can decrease the computational cost [65-67]. Our proposed model is also a shallow network that obtained higher accuracy. In fact, the lower number of layers requires a lower number of training parameters which can decrease training time and computational cost.

5) Using the dropout technique in the proposed model not only prevented overfitting but also increased its generalization and robustness.

6) The proposed model has more trainable parameters compared to other models, so it needs more time to train. Also, increasing the number of samples in the minority class increases the training time.

## 5. CONCLUSION

Accurate prediction of migration candidate virtual machines can reduce energy consumption in cloud data centers which can be possible by classifying the virtual machine in the workload class. In this paper, the combination of CNN and GRU was utilized to classify Microsoft Azure virtual machines. Due to the fact that collecting real data of virtual machine workload may lead to data imbalance, SMOTE method was used in our proposed model to overcome this issue. This method can improve the classification performance by increasing the number of training samples. Based on the result of the experiment, the proposed model presented higher accuracy compared to other existing state of the arts.

As mentioned, the use of the proposed model increases training time due to more trainable parameters and increasing the number of samples. To overcome this challenge, Google Colab cloud service has been usedWith access to free GPUs and TPUs Deep learning models can be trained in minutes or seconds.

The Microsoft Azure virtual machines dataset that was used in our experiment was collected in 2017. Using new datasets and other combinations of deep neural networks can be also worth exploring. Generative Adversarial Networks (GAN) can be also used to generate artificial samples. The proposed model can be employed to select the appropriate servers and reduce energy consumption. The proposed model can be leveraged in other practical applications, such as cloud data centers security, user request management, and resource utilization prediction.